\documentclass{article}
\usepackage[utf8]{inputenc}

\title{}
\usepackage[utf8]{inputenc}
\usepackage{amsmath}
\usepackage{amsfonts}
\usepackage{amssymb}
\usepackage[english]{babel}

\usepackage{color}
\usepackage{epsfig}
\usepackage{cite}
\usepackage{graphicx}%
\usepackage{dcolumn}%
\usepackage{bm}%

\begin{document}
\begin{center}
{\large\bf Measurement of Hubble Constant: Do Different Cosmological Probes Provide Different Values?} \\
\vspace*{0.5cm}
Rahul Kumar Thakur\textsuperscript{a,}\footnote{p20160425@hyderabad.bits-pilani.ac.in}, Shashikant Gupta \textsuperscript{b}\footnote{shashikantgupta.astro@gmail.com}, Rahul Nigam\textsuperscript{a,}\footnote{rahul.nigam@hyderabad.bits-pilani.ac.in}, PK Thiruvikraman \textsuperscript{a,}\footnote{thiru@hyderabad.bits-pilani.ac.in}\\
\vspace*{0.5cm}
$^a$ Birla Institute of Technology $\&$ Science, Pilani- Hyderabad Campus,\\Hyderabad, India.\\
$^b$ G D Goenka University,\\ Gurgaon, Haryana, 122103, India\\

\vspace*{0.1cm}
\end{center}
\begin {abstract}
Different measurements of the Hubble constant ($H_{0}$) are not consistent and a tension between the CMB based methods and cosmic distance ladder based methods has been observed. Measurements from various distance based methods are also inconsistent. To aggravate the problem, same cosmological probe (Type Ia SNe for instance) calibrated through different methods also provide different value of $H_{0}$. 

We compare various distance ladder based methods through the already available unique data obtained from Hubble Space Telescope (HST).
Our analysis is based on parametric (T-test) as well as non-parametric statistical methods such as the Mann-Whitney U test and Kolmogorov-Smirnov test. Our results show that different methods provide different values of $H_0$ and the differences are statistically significant. The biases in the calibration would not account for these differences as the data has been taken from a single telescope with common calibration scheme. The unknown physical effects or issues with the empirical relations of distance measurement from different probes could give rise to these differences.

\end{abstract}
\newpage

\maketitle
\section{Introduction}
\label{sec:intro}
Our understanding about the Universe has increased exponentially during the last century. Initially, static models were suggested but after Edwin Hubble's observations it was established that the Universe is expanding \cite{hubble1929,Freedman2000,Bieri2009}. These observations can be summarised in the form of a graph between the distance to several galaxies and their recessional velocities. The graph, known as the Hubble diagram, results in a straight line whose slope is called the Hubble parameter  \cite{Bergh1995}.The Hubble parameter is dynamic in nature and its present value is referred to as the Hubble Constant ($H_0$). It is among the most fundamental parameters of standard cosmology, as it is a measure of the expansion rate and age of the universe. The Hubble constant also determines the critical density, $\rho_c = 3H_0/8\pi G$, required for flat geometry of the Universe. Many other cosmological parameters such as physical properties of galaxies and quasars, growth of large scale structures, etc, depend on the numerical value of the Hubble constant. Thus, measuring the accurate value of $H_0$ is one of the most important challenges in present day cosmology. 

Various fundamentally different methods have been employed to determine the accurate value of Hubble constant including Cosmic Microwave Background Radiation (CMBR), gravitational waves, and the methods based on cosmic distance ladder such as Type Ia Supernovae (these methods are discussed in  \ref{sec:h0methods}). The latest CMBR observations by Planck satellite \cite{plank2013,Aghanim2018} along with the $\Lambda CDM$ cosmological model predicted $H_0=66.93\pm 0.62$ km/S/Mpc. It differs by around $3 \sigma$ or more with that measured with the distance ladder based methods \cite{Riess2016,Riess2019}.
Several solutions have been suggested including a dark component in the early universe \cite{Slatyer2018} to resolve this tension. It is also a matter of concern that various methods based on cosmic distance ladder too provide different values of $H_0$. Moreover, inconsistency has been observed in the data obtained from same secondary distance indicator (Type Ia Supernovae) calibrated through different probes (Cepheids vs Tip of Red Giant Branch) \cite{Freedman20,Freedman19,Yuan19}.
We wish to explore the statistical significance of the differences in these distance ladder based measurements alone. Significant differences would indicate either the lack of understanding of physical concepts used in these probes, or presence of systematic effects, or  issues related to the calibration of telescopes.

This paper is organised as follows: we discuss various distance ladder based methods to measure the Hubble constant 
in section \ref{sec:h0methods}. The data and statistical tools used for our analysis have been discussed in section \ref{sec:hst} and \ref{sec:methodology} respectively. The results and conclusions have been presented in section \ref{sec:result} and \ref{sec:conclusion}.

\section{Measurement of Hubble constant and cosmic distance ladder}
\label{sec:h0methods}
Determination of the Hubble constant requires measurement of the distances to the galaxies and their recessional velocities up to sufficiently large scales. The progress in accurate determination of $H_0$ has been slow due to various issues in distance measurement methods. A distance indicator, in principle, should fulfill the following basic criteria: (i) It should be bright enough to be detectable at
cosmological distances, (ii) the physics of the distance indicator should be well understood (iii) empirical relations among various quantities which are used for distance measurement should be free of systematic effects and (iv) statistically significant samples of such objects should be available. 

The period luminosity (P-L) relation of young Cepheid stars found in spiral galaxies is a promising primary distance indicator as the physics of this correlation is quite well understood \cite{lanoix1999}.  However, they are found in dusty regions and hence the observed P-L relation often requires correction against scattering, absorption, reddening and extinction leading to systematic errors. The P-L relation may also depend on the chemical composition \cite{Freedman1990} which is difficult to model. It is also difficult to resolve the Cepheid stars in distant galaxies, moreover, they are often observed in spiral galaxies. Thus the Cepheid variables alone can not be used to estimate $H_0$ and we need to use secondary distance measurement methods such as type Ia supernovae and Tully-Fisher relation. However, measuring accurate extra-galactic distances has always been challenging; often the uncertainties in measurement are underestimated and systematic effects dominate. Even today, identifying and reducing the sources of systematic errors in distance measurement is a challenging task. Measurement of $H_0$ up to 1 \% accuracy is clearly a difficult goal. However, as per the results of Hubble Space Telescope (HST) key project \cite{Freedman2001}, an accuracy of $H_0$ to 10\% was achieved in 2003; and 2.4\% accuracy has been claimed recently \cite{Riess2016}.Following different methods were used as secondary indicators: (i) Type Ia Supernovae (SNe Ia),(ii) Tully Fisher (T-F) relation, (iii) Surface Brightness Fluctuations, (iv) Type II Supernovae and (v) Fundamental Plane relation of elliptical galaxies. Below we review the current status of these methods.
\subsection{Type Ia Supernovae}
Type Ia supernovae (SNe Ia) are believed to arise from the explosion of a carbon-oxygen white dwarf \cite{Nomoto1997}. Their peak luminosity can outshine the entire host galaxy which makes them observable at cosmological distances. 
The peak luminosity of all SNe Ia are found to be in a narrow range. Moreover, the decline rate of their brightness is strongly correlated with the peak luminosity \cite{PHILLIPS1993}. Using the correlation, the absolute magnitude of an individual SN and hence its distance can be measured with high precision (less than 10\%). Thus, so far, SNe Ia are the most promising cosmological distance indicators \cite{Nomoto1997}. Unfortunately, the exact mechanism of the SN Ia explosion has not yet been well understood and recently, subclasses within the SNe Ia family have also been explored \cite{Umeda1999}. Some peculiar SNe Ia follow a special category recently termed as SNe Iax \cite{Jha2017}.   
Confidence in this empirical method will be strengthened once we gain theoretical understanding of the explosion process.

\subsection{The Tully–Fisher relation}
An empirical relation, known as Tully-Fisher relation, exists between the total luminosity and rotation speed of spiral galaxies \cite{Tully1997}. The correlation becomes tighter at longer wavelengths, especially, at I band the rms scatter is only 0.4mag \cite{willick1997}. The Tully-Fisher relation reflects the fact that massive and hence luminous galaxies rotate more rapidly; and can be used to measure the distances to spiral galaxies. Although role of dark matter is not certain and there are claims that the T-F relation is fundamentally a relation between rotation speed and total baryonic mass \cite{Gurovich}. However, the relation has been measured for hundreds of galaxies and amounts to around 15\% uncertainty in distance measurement \cite{Giovanelli1997}.

\subsection{Fundamental Plane Relation}
Correlation between luminosity and central velocity of elliptical galaxies was first discovered by \cite{Faber1976}. 
It is similar to T-F relation of spiral galaxies, however, it has lot of scatter. A similar noisy correlation also exists between effective radius and mean surface brightness of elliptical galaxies \cite{Davis1987}.
Both these correlations are now understood as the projections of a plane, known as the fundamental plane of elliptical galaxies, defined as $r_e \propto \sigma^{\alpha} I_e^{\beta} $, where, $r_e$ is the effective radius, $I_e$ is the average surface brightness within $r_e$ and $\sigma$ is the stellar velocity dispersion \cite{Schaeffer1993}. The scatter in the fundamental plane relation is much smaller than its projections and thus can be used to measure the distance to elliptical galaxies in various clusters

\subsection{Surface Brightness fluctuations}
Since each galaxy contains a finite number of stars, the number of stars in a small patch of a galaxy varies from point to point. This leads to fluctuations
in the surface brightness of the galaxy. These fluctuations smooth out with distance since the resolution of stars within the galaxies depends on distance \cite{Tonry2000}.  
This method is appropriate for elliptical galaxies because they have fairly consistent stellar populations or to spirals with prominent bulges. However, corrections due to variations in metallicity and age of galaxies are often required. Stellar population modeling and Cepheid variables are used for calibration of the method. 

\subsection{Type II Supernovae}
Type II SNe originate from core collapse of massive stars and are fainter than SNe Ia. Often they are observed in spiral arms of galaxies and HI clouds in Interstellar medium, but rarely in elliptical galaxies. Although, they are not standard candles, their distance can be measured by combining the spectra of expanding photosphere and photometric observations of angular size. This technique is known as Baade-Wesselink method \cite{hamuy2002}.  

\section{HST Key Project and HST Data}
\label{sec:hst}
The data for our analysis has been taken from Hubble Space Telescope Key (HST Key) Project \cite{Freedman2001}. The main goal of the HST Key project was to determine $H_0$ to an accuracy of $<10\%$ by calibrating the secondary distance indicators using Cepheid variables. Several new Cepheid stars were discovered by HST in various galaxies within $25$ Mpc distance. The Cepheid P-L relation was also calibrated against the metalicity by HST. The better seeing conditions and the ability to schedule the observations independent of phases of moon or weather conditions were the biggest advantage of HST compared to ground based observatories. Due to this, the number of Cepheids available for calibration increased drastically which is responsible for reduced uncertainties in the distance measurement. Table 1 of \cite{Freedman2001} compares the status of Cepheid calibrators pre and post HST.

\subsection{The revised P-L relation of Cepheid variables}
The Cepheid P-L relation was first introduced by Leavitt \cite{Leavitt1912}. 
Various authors e.g., \cite{Gieren1993}, used Cepheid surface brightness to estimate distances and absolute magnitudes.
Currently, Cepheids are  among the best stellar distance indicators and an important initial step on the cosmic distance ladder. However, the sensitivity of the zero point of P-L relation on the chemical composition has always been a matter of concern. 

Measuring an accurate value of $H_{0}$ was one of the motivating reasons for building the Hubble Space Telescope (HST). Thus, in the mid 1980s, accurate measurement of $H_0$ with an accuracy of 10\%, by observing several Cepheids and hence calibrating the secondary distance indicators was designated as one of the Key Projects of the HST. Before the launch of HST, most Cepheid searches were restricted to our own local Group of galaxies and the very nearest surrounding groups (M101, Sculptor, and M81 groups) \cite{Freedman1991}. By that time, only five  galaxies with well-measured Cepheid distances were available for calibration of the Tully-Fisher relation. While the calibration of the surface brightness fluctuation method \cite{Tonry1991} was done by using a single Cepheid distance, i.e., M31. Moreover, before HST no Cepheid calibrators were available for Type Ia supernovae. A large number of Cepheid variables were required to be observed to improve the calibration of P-L relation and hence the calibration of the secondary distance indicators. Several improvements and refinements were made by HST team including installation of HST Wide Field and Planetary Camera 2 (WFPC2) for photometric calibration. Observations of several Cepheids in the Large Magellanic Cloud (LMC) were carried out  define  the  fiducial P-L relation and to study the dependence of P-L relation on metallicity. The final results of HST key project were based on a Cepheid calibration of several secondary distance methods applied up to a distance $400$ Mpc.

\subsection{Data}
\label{sec:data}
Calibration is often a challenging issue in the astronomical measurements. Different primary indicators (such as Cepheids, TRGB, etc.) and different instruments are used for photometric calibration, which lead to systematic biases. 
The HST key project data is unique as it has been obtained through calibration of several different methods by a single primary distance indicator, i.e., Cepheid variables. Use of a single instrument (WFPC2 of HST) for photometric calibration also makes it special. 

Based on the revised Cepheid P-L relation, 36 type Ia SNe were calibrated which are available in table 6 of the Key Paper \cite{Freedman2001}. The value of $H_0$ obtained from this sample, $71\pm 2 \pm 6$ km/S/Mpc, is slightly higher than the previous measurement using SNe Ia ($68\pm 2 \pm 5$km/S/Mpc. Twenty one galaxies in general field and in various clusters and groups were calibrated for T-F relation using the newly available Cepheids. The measurements are available in table 7 of the key paper and provide a value of $H_0 = 71\pm 3 \pm 7$ km/S/Mpc. These results have not changed much from the previous measurements available in the literature \cite{sakai} 
indicating the self-consistency of T-F relation with respect to the Cepheid P-L relation. Distances to 11 elliptical galaxies in various clusters were measured through the Fundamental plane relation with a calibration using revised P-L relation. The new value of $H_0$ is $82\pm 6 \pm 9$ km/S/Mpc is substantially different from the previous measurements. The reason being, the galaxies in the key project were quite distant and their metalicities were quite high. Thus, the new calibration had larger impact on this sample. The other two methods, surface brightness fluctuations and type II SNe were also calibrated and used for $H_0$ measurement. However, the sample size of these methods are very small and hence we do not use them in our analysis. Altogether, we have three samples for our analysis, 36 SNe Ia, 21 T-F galaxies and 11 FPR galaxies. 

\section{Methodology}
\label{sec:methodology}
Five different probes have been employed to measure the Hubble constant in the HST key project. All the methods have been calibrated using a common mechanism, i.e., the P-L relation of Cepheid variables. However, empirical relations and the physical concepts involved in these methods are different. It'd be interesting to ask the following question: Do these methods provide same value of $H_0$. If the answer is no, it raises doubts about our understanding of the physical concepts and the empirical relations involved. Alternatively, it could also be due to the inconsistency of the systematics involved in the measurement process. Various statistical methods, related to hypothesis testing, can be employed in order to answer the above question. If $\mu_1$ and $\mu_2$ are the averages of $H_0$ from two different samples, then the null hypothesis can be set as-
\begin{equation}
    \mu_1 = \mu_2 .
    \label{eq:null}
\end{equation}
The alternative hypothesis which does not emphasise any of the samples, known as non-directional hypothesis, would be- 
\begin{equation}
    \mu_1 \neq \mu_2 ,
    \label{eq:alt-1}
\end{equation}
The alternative hypothesis would be directional if it emphasises on a particular sample, i.e., 
\begin{flalign}
\emph{$\mu_1 < \mu_2$ or $\mu_1 > \mu_2$}.  
\label{eq:alt-2}
\end{flalign}

Various parametric methods such as T-test are often employed for hypothesis testing when the data samples are drawn from the Gaussian distribution. As, the central limit theorem suggests that outcomes of a measurement process would follow Gaussian distribution, we expect the same for our data sets as well. However, non-parametric methods provide more reliable results when the distribution of data values is far from the Bell shape curve. We briefly discuss these techniques in the next section.

\subsection{Parametric Methods: T-test}
To test the Null hypothesis the average values of $H_0$ obtained from two different methods can be compared. The T-test based on the difference of means in terms of standard deviations which is known as T-score is often used. Since, the number of data points are different for different methods, unpaired T-test would be a suitable choice. If mean values of $H_0$ of two different samples with $n_1$ and $n_2$ data points are $M_1$ and $M_2$ respectively then the T-score is defined as
\begin{equation}
T = \frac{\mu_{1}-\mu_{2}}{\sqrt{\frac{s_{1}^2}{n_{1}}+ \frac{s_{2}^2}{n_{2}}}} ;
\label{eq:t-test}
\end{equation}
where $s_1^2$ and $s_2^2$ are the variance of first and second sample respectively\cite{George2003}. The uncertainties are important part of the measurement process and contain vital information. We thus weigh the measurements, $H_0$ values, with the uncertainties; so that more precise values get more weight. 
\begin{equation}
H_0^{'i} = \frac{{H_0^i}/{\sigma_{i}^2}}{\sum_j 1/\sigma_j^2}
\label{Eq:WMean}
\end{equation}
where $H_0^{i}$ represents the individual measurement of Hubble constant \cite{Ratra2015, Podariu}.
If the alternative hypothesis is directional, i.e., $M_1> M_2$ or vice-versa (as shown in Eq~\ref{eq:alt-2}), only one tail of the distribution is used for computation of probability and is known as ``one tailed test". Since, our alternative hypothesis in Eq~\ref{eq:alt-1} is non-directional, as it does not emphasise on any particular method ($\mu_1 \neq \mu_2$), a ``two tailed test" would be required. Large value of T-score in Eq~\ref{eq:t-test} indicates that the means of the two samples are quite different. In order to test if the difference is significant one needs to compare the T score calculated from Eq~\ref{eq:t-test} with a critical value which is the probability of obtaining the data samples assuming that the Null hypothesis is true. It is often available in the tables and depends on degrees of freedom, i.e., the number of data points in different samples. If the T-score is larger than the critical value the null hypothesis is rejected while a smaller T-score supports the null hypothesis. 
\\

\subsection{Non-parametric Methods}
The underlying assumption while performing T-test is that the variables ($H_0$ values in our case) follow normal distribution. However, often due to systematic effects either in the measurement process or in the instruments, the data values are far from normal distribution. In such cases we need to apply the statistical tests which do not demand the assumptions about the population parameters from which the sample is drawn. These are known as distribution free tests or non-parametric tests. Below we outline some non-parametric methods which can be applied on the data \cite{George2003}. 

\subsubsection{Mann-Whitney U test}
\label{sec:u-test}
Due to its resistance to the outliers, median is a more robust estimate of central tendency than mean. The Mann-Whitney U test or rank sum test is a non-parametric alternative to the un-paired T-test. It compares the median of two samples to test if the samples are drawn from same population. If the data values follow normal distribution, the relative efficiency of parametric methods is higher and vice-versa. The null hypothesis in this case would be-
\begin{equation}
    M_1 = M_2 ,
    \label{eq:null-2}
\end{equation}
where $M_1$ and $M_2$ are the medians of the samples respectively. The non-directional alternative hypothesis would be $M_1 \neq M_2$ and the directional hypothesis would be either $M_1 < M_2 $ or $M_1 > M_2 $. 

In the Mann-Whitney U test, the data values of the two samples are arranged in ascending order and ranks are assigned to the combined data. The smallest data value gets a rank $1$. Now the sum of ranks of each sample ($\Sigma R_1$ and $\Sigma R_2$, say) are calculated. The $U$ values are computed using Eq~\ref{eq:u1} and \ref{eq:u2}. 
\begin{equation}
U_{1} = n_{1}n_{2}+\frac{n_{1}(n_{1}+1)}{2}-\sum R_{1}
\label{eq:u1}
\end{equation}
\begin{equation}
U_{2} = n_{1}n_{2}+\frac{n_{2}(n_{2}+1)}{2}-\sum R_{2}
\label{eq:u2}
\end{equation}
where $n_1$ and $n_2$ are the sample sizes. One can easily verify that both $U_1$ and $U_2$ are always positive and $U_1 +U_2 = n_1 n_2$. The smaller value of $U_1$ and $U_2$ is designated as the $U$ statistic. Now statistical tables are used to calculate the critical value of U statistic for a given significance level. A comparison of the critical $U$ value with the obtained $U$ decides the rejection or non-rejection of the null hypothesis. Tables for critical $U$ value are available only for small sample sizes and for large samples approximate normal deviate $z$ is calculated \cite{George2003}
\begin{equation}
z = \frac{(|U-\frac{n_{1}n_{2}}{2}|-0.5)}{A} ,
\label{eq:z}
\end{equation}
where $A = \sqrt\frac{n_1{n_2}(n_1+n_2+1)}{12}$. Mode in the numerator signifies the fact that $(U- n_1n_2)/2$ is always negative. Tables of normal distribution can now be used to calculate the critical $z$ for a given significance level. If the $z$ calculated through Eq~\ref{eq:z} is greater than or equal to the critical $z$ the null hypothesis is rejected.

\subsubsection{Kolmogorov-Smirnov Test}
\label{sec:kstest}
The Kolmogorov Smirnov test (K-S test) is a non-parametric test of equality which can be used to compare a data sample with a reference probability distribution \cite{George2003}. Alternatively, it can also be used to compare two different samples by computing their cumulative distribution functions (CDF). It quantifies a distance between the empirical distribution functions of both the samples. The distributions are calculated under the null hypothesis that both the samples are drawn from the identical distribution. The KS test is quite useful since it is sensitive to the difference in location as well as shape of the underlying distributions.

The two-sample KS test estimates the difference between the CDFs of the two sample data vectors over a given range of $x$ in each data set. The test statistic in the two-sided test is the maximum distance, $D$, between the CDFs of the distributions of the two data vectors:
\begin{equation}
D = max |F_1(x)-F_2(x)| \,    
\label{eq:dist}
\end{equation}
where $F_1(x)$ and $F_2(x)$ are the proportions of $x_1$ and $x_2$ values that are less than or equal to $x$.
For our analysis we have have applied KS test function, available in Matlab, $h = kstest2(x_1,x_2)$, where $x_1$ and $x_2$ are the two samples. Based on the maximum distance, $D$, between the CDFs calculated using $x_1$ and $x_2$, it returns a value of zero or one. For large value of $D$ the function returns $h = 1$ which implies rejection of the null hypothesis at the  given significance level, $\alpha$. For small value of $D$ the function returns  $h = 0$, and this implies a failure to rejection of the null hypothesis at the given significance level, which is $\alpha = 0.01$ in our analysis.

\section{Results and Discussion}
\label{sec:result}
We first calculate the average value of Hubble constant and its standard deviation obtained from each method and present it in table-\ref{Table:MeanH0}. It is clear that although the average value of $H_0$ are close to each other for SNe and TF but their standard deviations are very different. The average in case of FPR is quite different from the other two methods. 

\begin{figure}
\centering
\includegraphics[width=14.0cm]{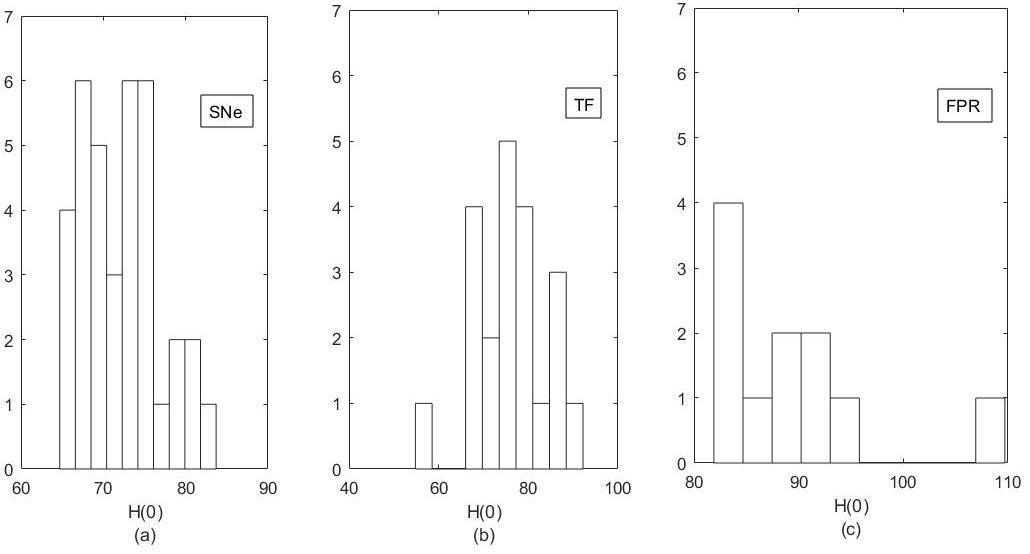}
\caption{Frequency distribution of $H_0$ values obtained from different methods: (a) SNe Ia (b) Tully Fisher (c) Fundamental Plane Relation}
\label{fig:Hist}
\end{figure}

Now, we sketch histogram of the $H_0$ values in the three samples which are presented in figure~\ref{fig:Hist}. A first glance at the figures indicates a deviation from normal distribution.
Although, small sample size in case of Tully-Fisher and Fundamental Plane relation could be one of the reasons for the deviation, the SNe sample size is sufficient to have shown the Bell shape. In any case, this motivates us to use the non-parametric methods. Thus, along with the regular parametric tests we apply non-parametric methods as well. 

\begin{center}
\begin{table}
\centering
\begin{tabular}{ |c|c|c|c|c|} 
 \hline
S N  & Method & Mean $H_0$ & Std dev & No of data points \\ 
 \hline
 1 & SNe & 72.18 & 4.87 & 36\\
 \hline
 2 & TF & 75.68 & 8.34 & 21\\
 \hline
 3 & FPR & 89.3 & 8.07 & 11\\
 \hline
\end{tabular}
\caption{Frequency distribution of $H_0$ values obtained from different methods: (a) SNe Ia (b) Tully Fisher (c) Fundamental Plane Relation.}
\label{Table:MeanH0}
\end{table}
\end{center}
 
\subsection{Results for parametric test: T-test}
As mentioned in \ref{sec:data} there are three samples of data and hence, three pairs of data samples namely, pair 1: SNe-TF, pair 2: TF-FPR and pair 3: SNe-FPR. Our first pair of samples consists of data from Type Ia SNe and Tully Fisher methods. T-test is performed to test the null hypothesis (\ref{eq:null}) that the average value of $H_0$ is same in SNe and T-F methods. The T-score has been calculated using Eq~\ref{eq:t-test} and is presented in table-\ref{table:t-test-1} (row 1). The critical value for the non-directional alternative hypothesis (\ref{eq:alt-1}) : ``average value of $H_0$ for the two methods are not same" taken from the table at 99\% significance level is  2.7. Since the observed T value is smaller than the critical value, the null hypothesis is not rejected. However, one tailed T test is implemented , if the alternative hypothesis is directional, i.e., $H_0(SNe)$ is smaller than $H_0(TF)$. The critical value at 95\% significance level  is $1.7$ which is smaller than the T-score and the Null hypothesis is rejected. Similar analysis for the other two pairs is also presented in table-\ref{table:t-test-1}. It is clear that for both pairs the obtained value is greater than the critical value, hence, the Null hypothesis is rejected  at 99\%. 

We now include the measurement uncertainties in the analysis by defining a new variable $H_{0}^{'}$ as the $H_0$ values weighted by the measurement uncertainties as defined in Eq~\ref{Eq:WMean} in each sample. The T-score for various pairs of samples are again calculated and are presented in table~\ref{table:t-test-2}. Clearly the T-score is larger compared to the critical value in all cases. Hence the null hypothesis ``the measurement values are same using different methods" is rejected in all cases. 

\begin{center}
\begin{table}
\centering
\begin{tabular}{ |c|c|c|c|c|c|c|c|c| } 
 \hline
 Pair &$\mu_{1}$  & $\mu_{2}$ & $S_{1}^{2}$ & $S_{2}^{2}$  & $T_{theory}$ & $T_{critical}$ & $DOF$ & Result \\ 
 \hline
 SN-TF & 72.2 & 75.7 & 23.8 & 69.6  &  1.8 & 2.7 & $55$ & NR\\ 
 \hline
 TF-FPR & 75.7 & 89.3 & 69.6 & 65.2  &  4.5 & 2.8 & $30$ & R\\
 \hline
 SN-FPR & 72.2 & 89.3 & 23.8 & 65.2  &  6.7 & 2.7 & $45$ & R\\ 
 \hline
\end{tabular}
\caption{Unpaired T test between various pairs of data samples without using uncertainties. Null hypothesis (Eq~\ref{eq:null}) is rejected at 99\% significance level except SNe-TF pair.}
\label{table:t-test-1}
\end{table}
\end{center}

\begin{center}
\begin{table}
\centering
\begin{tabular}{ |c|c|c|c|c|c|c|c| c| } 
\hline
Pair & $\mu_{1}$ & $\mu_{2}$ & $S_{1}^{2}$ & $S_{2}^{2}$  & $T_{theory}$ & $T_{critical}$ & $DOF$ & Result \\ 
\hline
SN-TF & 2.0 & 3.5 & 0.2 & 0.3  &  11.9 & 2.7 & $55$ & R\\
\hline
TF-FPR & 3.5 & 8.0 & 0.3 & 7.1  &  5.5 & 2.8 & $30$ & R\\
\hline
SN-FPR & 2.0 & 8.0 & 0.2 & 7.1  &  7.5 & 2.7 & $45$ & R \\
\hline
\end{tabular}
\caption{Unpaired t test between various pairs of samples ($H_0$ values weighted by uncertainties). Null hypothesis (Eq~\ref{eq:null}) is rejected at 99\% significance level in all cases.}
\label{table:t-test-2}
\end{table}
\end{center}

\subsection{Results for Non-parametric tests}
Since, the histograms of data samples in fig~\ref{fig:Hist} do not show a clear bell shape we apply the non-parametric tests as well. The first is U test described in ~\ref{sec:u-test} and the next is KS test discussed in ~\ref{sec:kstest}.
\subsubsection{U test}
The null hypothesis for U test has been setup in Eq~\ref{eq:null-2} and the numerical values of $U_1$ and $U_2$ have been calculated using Eq~\ref{eq:u1} and \ref{eq:u2}. Since, the sample sizes are relatively large to obtain the critical $U$ from the table, the normal approximate $z$ has been calculated using Eq~\ref{eq:z}. All these values are presented in table~\ref{table:u-1}. The critical value of $z$ corresponding to the 99\% significance level is $2.58$ which is common in all cases. Since $z$ value in column $7$, obtained using Eq~\ref{eq:z}, is smaller for SN-TF pair, it supports the null hypothesis. On the other hand it is greater than $2.58$ for TF-FPR and SN-FPR pairs the null hypothesis is rejected for these sample pairs. However, the table value of $z$ at $95\%$ significance level is $1.96$ and thus the Null hypothesis is rejected for this case also at 95\% level.

In order to make use of the uncertainties in the measurement, we apply the U test on the $H_{0}^{'}$ values obtained using Eq~\ref{Eq:WMean} for all the three pairs of samples. As in the previous case, the $U_1$, $U_2$ and $z$ values have been calculated using Eqs~\ref{eq:u1}, \ref{eq:u2} and \ref{eq:z} and are presented in table~\ref{table:u-2}. It is clear from the table that since in all cases the tabled $z$ value is smaller than the calculated $z$ value, the null hypothesis is rejected at $99\%$ significance level in all cases. 

\begin{center}
\begin{table}
\centering
\begin{tabular}{ |c|c|c|c|c|c|c|c|c|} 
\hline
Pair & $n_{1}$  & $n_{2}$ & $U_1$ & $U_2$ & $U$ & $z$ from Eq~\ref{eq:z}  & $z$ from table & Result  \\  
\hline
SN-TF & 36 & 21 & 501.5 & 254.5 & 254.5 &2.03 & 2.58 & NR  \\
\hline
TF-FPR & 21 & 11 & 205 & 26 & 26 & 3.53  &  2.58 & R \\
\hline
SN-FPR & 36 & 11 & 392 & 4 & 4 & 4.86  & 2.58 & R  \\
\hline
\end{tabular}
\caption{U test between various pairs of samples (using original $H_0$ values). $z$ values in columns $8$ has been from table of standard normal distribution for $99\%$ significance level.  Null hypothesis (Eq~\ref{eq:null-2}) is rejected in two cases.}
\label{table:u-1}
\end{table}
\end{center}
\begin{center}
\begin{table}
\centering
\begin{tabular}{ |c|c|c|c|c|c|c|c|c|} 
\hline
Pair & $n_{1}$  & $n_{2}$ & $U_{1}$ & $U_{2}$ & $U$ & $z$ from Eq~\ref{eq:z}  & $z$ from table & Result  \\ 
\hline
SN-TF & 36 & 21 & 3 & 753 & 3 & 6.19 & 2.58 &  R  \\
\hline
TF-FPR & 21 & 11 & 1 & 230 & 1 & 4.5  & 2.58 &  R  \\
\hline
SN-FPR & 36 & 11 & 0 & 396 & 0 & 4.96  & 2.58 &  R   \\
\hline
\end{tabular}
\caption{U test between various pairs of samples ($H_0$ values weighted by uncertainties). $z$ values in columns $8$ has been from table of standard normal distribution for $99\%$ significance level. Null hypothesis (Eq~\ref{eq:null-2}) is rejected in all cases.}
\label{table:u-2}
\end{table}
\end{center}
\subsubsection{KS Test} 
\begin{figure}
    \centering
    \includegraphics[width= 15.0cm]{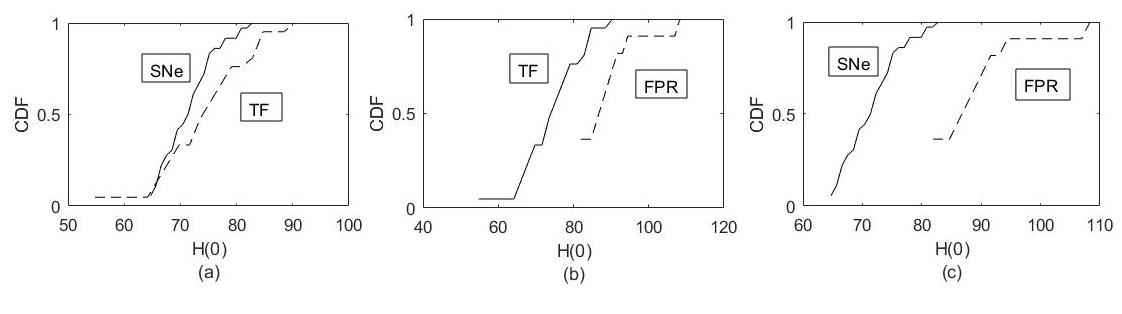}
\caption {K-S Test: Cumulative distribution function (CDF) for original values of Hubble constant obtained from different methods. CDF of different pairs have been plotted together for comparison: (a) SNe Ia and T-F Fig (b) T-F and FPR and (c) SNe Ia and FPR.}
\label{fig:ks-1}
\end{figure}

Finally, we perform the KS test for the same three pairs of samples to verify the equality of different methods (see \ref{sec:kstest}). The original $H_0$ values were supplied to the MATLAB function $kstest2$ to calculate the value of $h$ at $99\%$ significance level. The results are presented in table~\ref{table:ks-test}. The values of $h$ for SN-TF pair is zero which supports the Null hypothesis. However, in the remaining two cases $h$ is equal to one indicating the rejection of the null hypothesis. The CDFs of these pairs have been plotted in fig~\ref{fig:ks-1}. 
The CDFs obtained for TF-FPR pair and for SN-FPR pair are quite far from each other. Since, the distance between the CDFs is very large in both cases, the null hypothesis is rejected. Although, the CDFs for SN and TF do not match well, however, they are not too different. This is the reason for non rejection of the null hypothesis.

\begin{figure}
    \centering
    \includegraphics[width= 15.0cm]{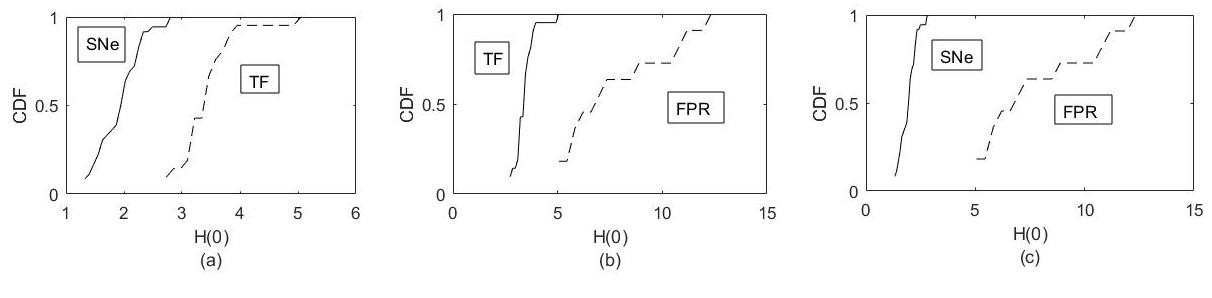}
\caption{K-S Test: Cumulative distribution function (CDF) for Hubble constant values weighted by uncertainties. Comparison of CDF of different samples: (a) SNe Ia and T-F Fig (b) T-F and FPR and (c) SNe Ia and FPR.}
\label{fig:ks-2}
\end{figure}

In order to make use of the information available in the uncertainties, we apply the KS test on the uncertainty weighted measurements, i.e., $H_{0}^{'}$. The results are presented in table~\ref{table:ks-test}. This time the null hypothesis is rejected in all the three cases. Thus we conclude that the different methods of measuring $H_0$ provide different results. The cumulative distributions of $H_0$ values from different methods have been compared in fig~\ref{fig:ks-2}. It is clear from the figure that in all the cases the distance between the cdfs is too large and hence the null hypothesis is rejected in all cases. 

\begin{center}
\begin{table}
\centering
KS test results \\
 \vspace{0.3cm}
 \centering
\begin{tabular}{ |c|c|c||c|c|c|c| } 
 \hline
 
 \hline
 Pair & h value & Result & Pair & h value & Result \\
 \hline
SN-TF  & 0 & NR & SN-TF & 1 & R \\ 
 \hline
 TF-FPR & 1 & R & TF-FPR & 1 & R \\ 
 \hline
 SN-FPR & 1 & R & SN-FPR & 1 & R \\ 
 \hline
 \end{tabular}
 \\ ($A$) \hspace{4.0cm}  ($B$)
\caption{$A$ shows the KS test results for samples without including uncertainties. Null hypothesis is rejected in two cases. $B$ shows KS test results for samples with including uncertainties. Null hypothesis is rejected in all cases.}
\label{table:ks-test}
\end{table}
\end{center}


\section{Conclusions}
\label{sec:conclusion}
We have applied both the parametric and non-parametric methods to test if measurement of $H_0$ from different methods based on cosmic distance ladder are statistically different. To avoid the systematic differences arising from instrumental bias we use the data from the same telescope, i.e., HST. Our null hypothesis is that the average or median $H_0$ values obtained from different methods are same. Based on the results presented in section \ref{sec:result} we conclude that the null hypothesis is rejected at 99\% confidence level in most cases and at 95\% level in all cases. A possible reason for the failure could be the gaps in our understanding of the physical processes involved in these cosmological probes. As an example, SNe Ia are assumed to be among the most precise secondary distance indicators. However, the possibility of sub-classes within SNe Ia class could bias the correlation between the peak luminosity and decline rate which might lead to undesired systematic effects in distance measurement. The physics of T-F relation is also poorly understood, since, the constant mass to light ratio assumption may not be reliable and the role of dark matter in galaxy rotation curve is also debatable \cite{Gurovich}. These issues could bias the distances measured through T-F relation. Difference in metallicity of various elliptical galaxies in FPR method could also affect the distance measurement. Although, the SNe Ia measurements have improved over time and precise measurement of $H_0$ using SNe Ia are available now, \cite{Riess2016} however, other methods have not been improved to the same level of sensitivity.

\section*{Acknowledgement} 
SG thanks SERB for financial assistance (EMR/2017/003714). RN, RKT and PKT extend their gratitude to BITS-Pilani Hyderabad Campus for providing all the required infrastructure. 




\begin{thebibliography}{1}
\bibitem{hubble1929} Hubble, Edwin. ``A relation between distance and radial velocity among extra-galactic nebulae." Proceedings of the National Academy of Sciences 15, no. 3 (1929): 168-173.
\bibitem{Freedman2000} Freedman, Wendy L. ``The Hubble constant and the expansion age of the Universe." Physics Reports 333 (2000): 13-31.
\bibitem{Bieri2009} Nussbaumer, Harry, and Lydia Bieri. ``Discovering the expanding universe." Discovering the Expanding Universe, by Harry Nussbaumer, Lydia Bieri, Foreword by Allan Sandage, Cambridge, UK: Cambridge University Press, 2009 (2009).
\bibitem{Bergh1995} van den Bergh, Sidney. ``A new method for the determination of the Hubble parameter." The Astrophysical Journal Letters 453, no. 2 (1995): L55.
\bibitem{plank2013} Hinshaw, Gary, D. Larson, Eiichiro Komatsu, David N. Spergel, CLaa Bennett, Joanna Dunkley, M. R. Nolta et al. ``Nine-year Wilkinson Microwave Anisotropy Probe (WMAP) observations: cosmological parameter results." The Astrophysical Journal Supplement Series 208, no. 2 (2013): 19.
\bibitem {Aghanim2018} Aghanim, N., Yashar Akrami, M. Ashdown, J. Aumont, C. Baccigalupi, M. Ballardini, A. J. Banday et al. ``Planck 2018 results. VI. Cosmological parameters." arXiv preprint arXiv:1807.06209 (2018).
\bibitem{Riess2016} Riess, Adam G., Lucas M. Macri, Samantha L. Hoffmann, Dan Scolnic, Stefano Casertano, Alexei V. Filippenko, Brad E. Tucker et al. ``A 2.4\% determination of the local value of the Hubble constant." The Astrophysical Journal 826, no. 1 (2016): 56.
\bibitem{Riess2019}Riess, Adam G., Stefano Casertano, Wenlong Yuan, Lucas M. Macri, and Dan Scolnic. ``Large Magellanic Cloud Cepheid standards provide a 1\% foundation for the determination of the Hubble constant and stronger evidence for physics beyond $\Lambda$CDM." The Astrophysical Journal 876, no. 1 (2019): 85.
\bibitem{Slatyer2018} Slatyer, Tracy R., and Chih-Liang Wu. ``Early-Universe constraints on dark matter-baryon scattering and their implications for a global 21 cm signal." Physical Review D 98, no. 2 (2018): 023013.
\bibitem{Freedman20}Freedman, Wendy L., Barry F. Madore, Taylor Hoyt, In Sung Jang, Rachael Beaton, Myung Gyoon Lee, Andrew Monson, Jill Neeley, and Jeffrey Rich. "Calibration of the Tip of the Red Giant Branch (TRGB)." arXiv preprint arXiv:2002.01550 (2020).
\bibitem{Freedman19}Freedman, Wendy L., Barry F. Madore, Dylan Hatt, Taylor J. Hoyt, In Sung Jang, Rachael L. Beaton, Christopher R. Burns et al. "The Carnegie-Chicago Hubble Program. VIII. An independent determination of the Hubble constant based on the tip of the red giant branch." The Astrophysical Journal 882, no. 1 (2019): 34.
\bibitem{Yuan19}Freedman, Wendy L., Barry F. Madore, Taylor Hoyt, In Sung Jang, Rachael Beaton, Myung Gyoon Lee, Andrew Monson, Jill Neeley, and Jeffrey Rich. "Calibration of the Tip of the Red Giant Branch (TRGB)." arXiv preprint arXiv:2002.01550 (2020).
\bibitem {lanoix1999} Lanoix, Pierre, Georges Paturel, and Robert Garnier. ``Bias in the Cepheid period-luminosity relation." The Astrophysical Journal 517, no. 1 (1999): 188.
\bibitem{Freedman1990} Freedman, Wendy L., and Barry F. Madore. ``An empirical test for the metallicity sensitivity of the Cepheid period-luminosity relation." The Astrophysical Journal 365 (1990): 186-194.
\bibitem{Freedman2001} Freedman, Wendy L., Barry F. Madore, Brad K. Gibson, Laura Ferrarese, Daniel D. Kelson, Shoko Sakai, Jeremy R. Mould et al. ``Final results from the Hubble Space Telescope key project to measure the Hubble constant." The Astrophysical Journal 553, no. 1 (2001): 47.

\bibitem{Nomoto1997} Nomoto, Ken’ichi, Koichi Iwamoto, and Nobuhiro Kishimoto. ``Type Ia supernovae: their origin and possible applications in cosmology." Science 276, no. 5317 (1997): 1378-1382..
\bibitem{PHILLIPS1993} Phillips, Mark M. "The absolute magnitudes of Type IA supernovae." The Astrophysical Journal 413 (1993): L105-L108.
\bibitem{Umeda1999} Umeda, Hideyuki, Ken'ichi Nomoto, Chiaki Kobayashi, Izumi Hachisu, and Mariko Kato. ``The origin of the diversity of type Ia supernovae and the environmental effects." The Astrophysical Journal Letters 522, no. 1 (1999): L43..

\bibitem{Jha2017} Jha, Saurabh W. ``Type Iax Supernovae." arXiv preprint arXiv:1707.01110 (2017).

\bibitem{Tully1997} Tully, R. Brent, and J. Richard Fisher. ``A new method of determining distances to galaxies." Astronomy and Astrophysics 54 (1977): 661-673.
\bibitem{willick1997} Willick, Jeffrey A., Stéphane Courteau, S. M. Faber, David Burstein, Avishai Dekel, and Michael A. Strauss. ``Homogeneous velocity-distance data for peculiar velocity analysis. iii. the mark iii catalog of galaxy peculiar velocities." The Astrophysical Journal Supplement Series 109, no. 2 (1997): 333.
\bibitem{Gurovich} Gurovich, Sebastián, Stacy S. McGaugh, Ken C. Freeman, Helmut Jerjen, Lister Staveley-Smith, and W. J. G. De Blok. ``The Baryonic Tully–Fisher Relation." Publications of the Astronomical Society of Australia 21, no. 4 (2004): 412-414.
\bibitem{Giovanelli1997} Giovanelli, Riccardo, Martha P. Haynes, Luiz N. da Costa, Wolfram Freudling, John J. Salzer, and Gary Wegner. ``The Tully-Fisher Relation and H0." The Astrophysical Journal Letters 477, no. 1 (1997): L1.
\bibitem{Faber1976}Faber, S. M., and Robert E. Jackson. ``Velocity dispersions and mass-to-light ratios for elliptical galaxies." The Astrophysical Journal 204 (1976): 668-683.
\bibitem{Davis1987} Djorgovski, S., and Marc Davis. ``Fundamental properties of elliptical galaxies." Astrophysical Journal 313 (1987): 59-68.
\bibitem{Schaeffer1993} Schaeffer, R., S. Maurogordato, A. Cappi, and F. Bernardeau. ``The fundamental plane of galaxy clusters." Monthly Notices of the Royal Astronomical Society 263, no. 1 (1993): L21-L26.

\bibitem{Tonry2000} Tonry, John L., John P. Blakeslee, Edward A. Ajhar, and Alan Dressler. ``The surface brightness fluctuation survey of galaxy distances. II. Local and large-scale flows." The Astrophysical Journal 530, no. 2 (2000): 625.
\bibitem{hamuy2002} Hamuy, Mario, and Philip A. Pinto. "Type II supernovae as standardized candles." The Astrophysical Journal Letters 566, no. 2 (2002): L63.

\bibitem{Leavitt1912} Leavitt, H. S., and E. C. Pickering. ``Harvard College Observ." Circ 173 (1912): 1.
\bibitem{Gieren1993} Gieren, Wolfgang P. ``Surface brightness distance determinations to the Large Magellanic Cloud Cepheid variables HV 899 and 2257." Monthly Notices of the Royal Astronomical Society 265, no. 1 (1993): 184-188.

\bibitem{Freedman1991} Madore, Barry F., and Wendy L. Freedman. ``The Cepheid distance scale." Publications of the Astronomical Society of the Pacific 103, no. 667 (1991): 933.
\bibitem{Tonry1991} Tonry, John L. ``Surface brightness fluctuations-A bridge from M31 to the Hubble constant." The Astrophysical Journal 373 (1991): L1-L4.
\bibitem{Gibson2000a}Gibson, Brad K., Philip R. Maloney, and Shoko Sakai. ``Has blending compromised cepheid-based determinations of the extragalactic distance scale?." The Astrophysical Journal Letters 530, no. 1 (2000): L5.
\bibitem{sakai} Mould, Jeremy R., Shaun MG Hughes, Peter B. Stetson, Brad K. Gibson, John P. Huchra, Wendy L. Freedman, Robert C. Kennicutt Jr et al. ``The Hubble Space Telescope Key Project on the Extragalactic Distance Scale. XXI. The Cepheid Distance to NGC 1425." The Astrophysical Journal 528, no. 2 (2000): 655.
\bibitem{George2003}Sheskin, David J. Handbook of parametric and nonparametric statistical procedures. Chapman and Hall/CRC, 2003.
\bibitem{Ratra2015} Crandall, Sara, and Bharat Ratra. ``Non-Gaussian error distributions of LMC distance moduli measurements." The Astrophysical Journal 815, no. 2 (2015): 87.
\bibitem{Podariu} Podariu, Silviu, Tarun Souradeep, J. Richard Gott III, Bharat Ratra, and Michael S. Vogeley. ``Binned cosmic microwave background anisotropy power spectra: Peak location." The Astrophysical Journal 559, no. 1 (2001): 9.

\end{thebibliography}
\end{document}